\documentclass[12pt]{article}
\usepackage[margin=1in]{geometry}
\usepackage{amsmath}
\usepackage{amsthm}
\usepackage{graphicx}
\usepackage[official]{eurosym}
\usepackage{url}
\usepackage{tablefootnote}
\usepackage{multirow}
\usepackage{subfig}
\usepackage{float}
\usepackage[hidelinks]{hyperref}
\hypersetup{
    colorlinks=true,
    linkcolor=blue,
    filecolor=magenta,      
    urlcolor=cyan,
}
\usepackage{caption}
\usepackage{booktabs}
\usepackage{float}
\providecommand{\keywords}[1]{\textbf{keywords ---} #1}

\title{Classification problem in liability insurance using machine learning models: a comparative study}

\begin{document}

\author{Marjan Qazvini \footnote{Correspondence to: Marjan Qazvini, Department of Actuarial Mathematics and Statistics, School of Mathematical and Computer Sciences, Heriot-Watt University, Dubai,
E-mail: M.Qazvini@hw.ac.uk, (marjan.qazvini@gmail.com) Phone Number: +971 (4) 5717000.} (April 2022)}
\date{}

\maketitle

\begin{abstract}
Underwriting is one of the important stages in an insurance company. The insurance company uses different factors to classify the policyholders. 
In this study, we apply several machine learning models such as nearest neighbour and logistic regression to the Actuarial Challenge dataset used by Qazvini (2019) to classify liability insurance policies into two groups: 1 - policies with claims and 2 - policies without claims.

\end{abstract}


\section{Introdoction}
The applications of Machine Learning (ML) models and Artificial Intelligence (AI) in areas such as medical diagnosis, economics, banking, fraud detection, agriculture, etc, have been known for quite a number of years. ML models have changed these industries remarkably. However, despite their high predictive power and their capability to identify nonlinear transformations and interactions between variables, they are slowly being introduced into the insurance industry and actuarial fields. There are two types of ML models: supervised learnings where the target \textit{attributes} or \textit{labels} are known and unsupervised learnings where the target attributes or labels are unknown. Further, we can categorise unsupervised ML models into clustering and association rules and supervised ML models into classification and regression models, where in the former, the target attribute is categorical, and in the latter, it is numerical. In the following, we review the literature that considers the applications of ML models and algorithms in insurance. Campbell (1986) applies cluster analysis to group car models. He considers a measure of dissimilarity based on a within-group weighted sum of squares. Carfora et al. (2018) apply cluster analysis to group drivers according to their driving habits. Cluster analysis has also been used in life insurance and for the valuation of annuity portfolios. See Gan and Valdez (2020) and references therein for a review of cluster analysis and its applications in actuarial science. W\"{u}thrich (2017) applies K-means clustering algorithms to classify drivers based on driving styles, such as speed and acceleration patterns extracted from telematics data. Gao and W\"{u}thrich (2018) replace the categorical classes produced by K-means algorithm with low-dimensional continuous features by employing dimensional reduction methods such as the Principal Component Analysis (PCA) and a feed-forward neural networks with one hidden layer. Then, Gao et al. (2019) investigate the predictive power of these models and conclude that they can provide better out-of-sample prediction for claims frequency than traditional models. Pesantes-Narvaez et al. (2019) compare XGBoost algorithm with logistic regression and find that in terms of out-of-sample predictability, simplicity and interpretability, logistic regression outperforms XGBoost algorithm in an unbalanced dataset. Henchaerts et al. (2020) compare regression trees, random forests and XGBoost with Generalised Linear Models (GLMs) and conclude that boosted trees perform better than GLMs. Li et al. (2018) combine random forest, PCA and nearest neighbour methods to propose an ensemble algorithm where the voting stage is replaced by the nearest neighbour algorithm. They use their model for automobile fraud detection and find that this model can improve out-of-sample prediction accuracy. 

\par In this paper, we consider the classification algorithms and models that can be used both with numerical and categorical attributes. Classification is a problem that we encounter in our daily lives. We need to assign different objects to different groups or \textit{classes}. In most cases, these classes are mutually exclusive. Therefore, our goal is to determine \textit{decision regions} which are bounded by decision boundaries or decision surfaces. These boundaries can be linear or nonlinear. In this paper, we consider a binary classification problem, i.e. we have only 2 classes: 1 - policies with claims and 2 - policies without claims. We apply several machine learning algorithms and models to a pricing game dataset \footnote {http://cas.uqam.ca.}. This dataset contains information for liability insurance policies with a lot of zero claims which is typical of this type of insurance. Qazvini (2019) considers this dataset and applies a Zero-inflated Poisson regression model to this unbalanced dataset to predict the number of claims. Here we adopt different preprocessing steps to Qazvini (2019) and apply different machine learning models to classify policies with and without claims and compare their accuracy with the traditional logistic regression model. The rest of the paper is organised as follows: in Section 2 we discuss preprocessing steps and analyse our dataset. In Section 3 we explain the classification algorithms and models that we use in this paper. In Section 4 we implement the models discussed in Section 3 and compare those models in terms of out-of-sample accuracy. In this paper, we use \textit{Scikit-learn}, a machine learning library and \textit{Mathplotlib}, a data visualisation library in Python. 

\section{Data analysis}
\begin{figure}
\centering
\includegraphics[scale = 0.6]{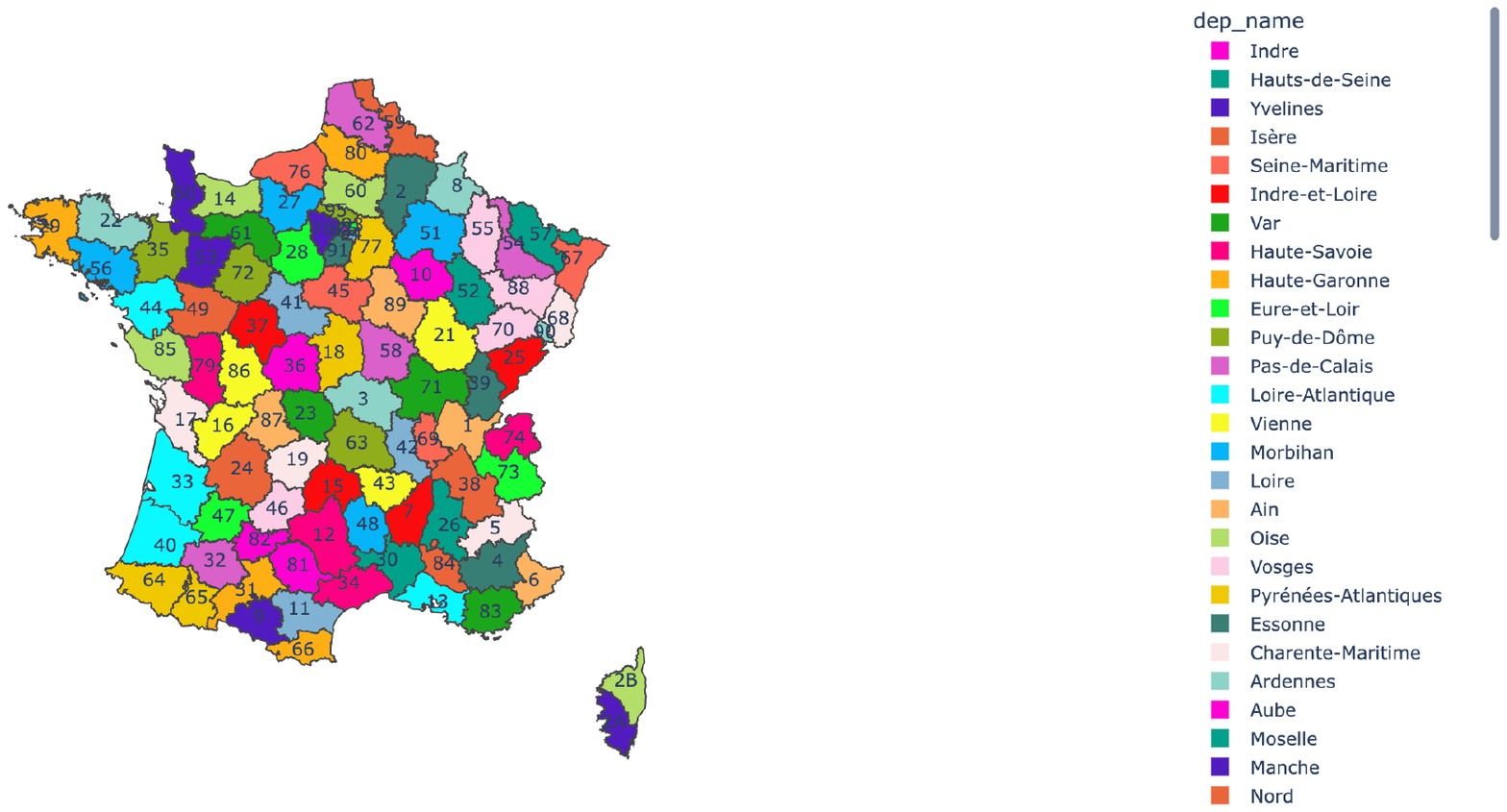}
\caption{\scriptsize Indre: 36, Hauts-de-Seine: 92, Yvelines: 78, Is\`{e}re: 38, Seine-Maritime: 76, Indre-et-Loire: 37, Var: 83, Haute-Savoie: 74, Haute-Garonne: 31, Eure-et-Loir: 28,
Puy-de-D\^{o}me: 63, Pas-de-Calais: 62, Loire-Atlantique: 44, Vienne: 86, Morbihan: 56, Loire: 42, Ain: 1, Oise: 60, Vosges: 88, Pyr\'{e}n\'{e}es-Atlantiques: 64, Essonne: 91, Charente-Maritime: 17,
Ardennes: 8, Aube: 10, Moselle: 57, Manche: 50, Nord: 59, Rh\^{o}ne: 69, Doubs: 25, Sa\^{o}ne-et-Loire: 71, Deux-S\`{e}vres: 79, Somme: 80, Seine-Saint-Denis: 93, Val-d`Oise: 95,
Ni\`{e}vre: 58, Landes: 40, Charente: 16, Marne: 51, Aude: 11, Alpes-Maritimes: 6, Vend\'{e}e: 85, Meuse: 55, Seine-et-Marne: 77, Jura: 39, Haut-Rhin: 68, C\^{o}tes-d'Armor: 22, Paris: 75,
Gard: 30, Corse-du-Sud: 2A, Dordogne: 24, Bas-Rhin: 67, Cantal: 15, Orne: 61, H\'{e}rault: 34, Finist\`{e}re: 29, Savoie: 73, Sarthe: 72, Meurthe-et-Moselle: 54, Gironde: 33, C\^{o}te-d'Or: 21, Eure: 27,
Val-de-Marne: 94, Haute-Vienne: 87, Galvados: 14, Haute-Sa\^{o}ne: 70, Cher: 18, Aisne: 2, Corr\`{e}ze: 19, Allier: 3, Tarn: 81, Haute-Marne: 52, Ari\`{e}ge: 9, Maine-et-Loire: 49, Loiret: 45, Ard\`{e}che: 7,
Creuse: 23, Aveyron: 12, Pyr\`{e}n\`{e}es-Orientales: 66, Lot-et-Garonne: 47, Ille-et-Vilaine: 35, Gers: 32, Bouches-du-Rh\^{o}ne: 13, Haute-Loire: 43, Loz\`{e}re: 48, Loir-et-Cher: 41, Yonne: 89, 
Haute-Corse: 2B, Lot: 46, Hautes-Pyr\'{e}n\'{e}es: 65, Alpes-de-Haute-Provence: 4, Hautes-Alpes: 5, Territoire de Belfort: 90, Tarn-et-Garonne: 82, Dr\^{o}me: 26, Mayenne: 53, Vaucluse: 84.}
\label{fig:Dep}
\end{figure}

\begin{figure}
\centering
\includegraphics[scale = 0.45]{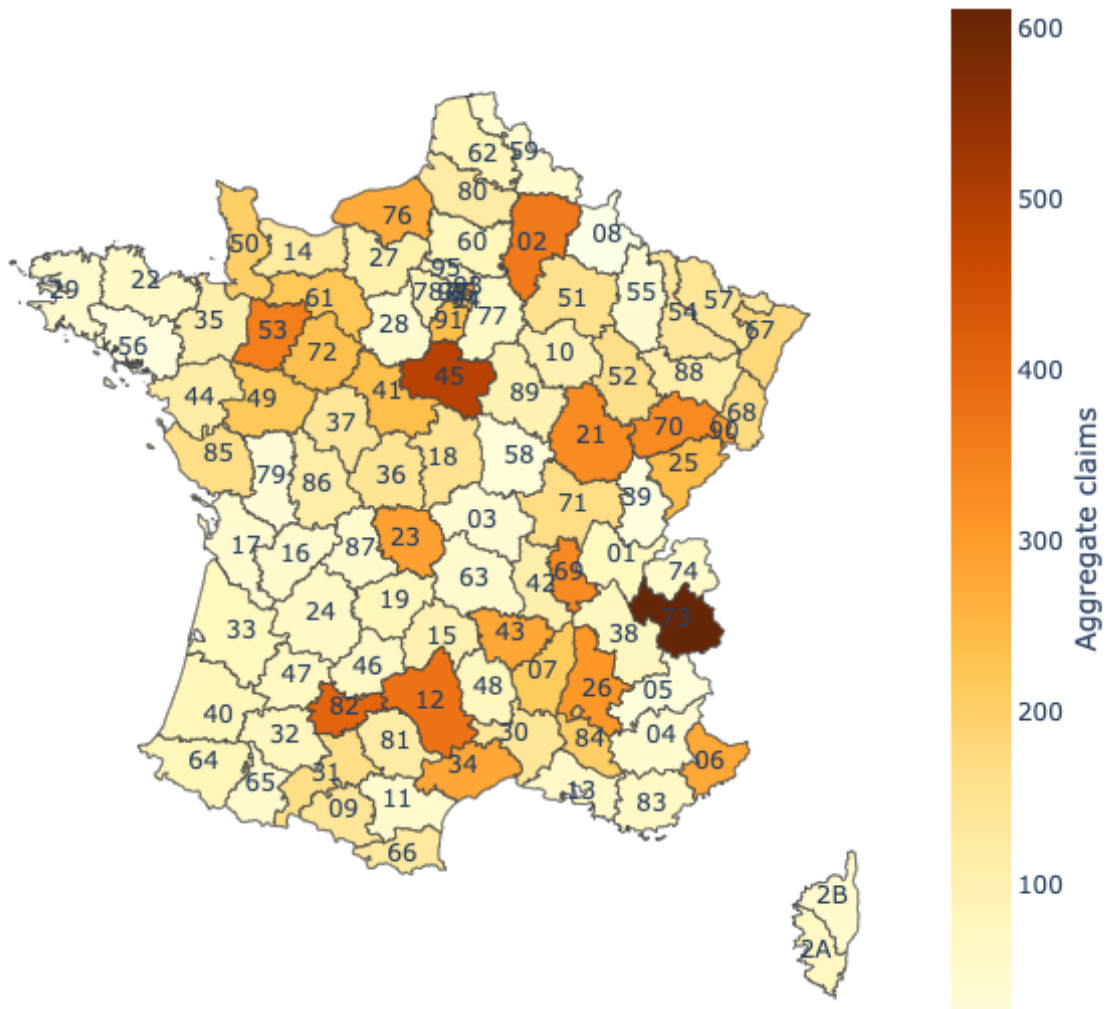}
\caption{\scriptsize The total number of claims per departments}
\label{fig:DepClaim}
\end{figure}

\begin{figure}
\centering
\includegraphics[scale = 0.45]{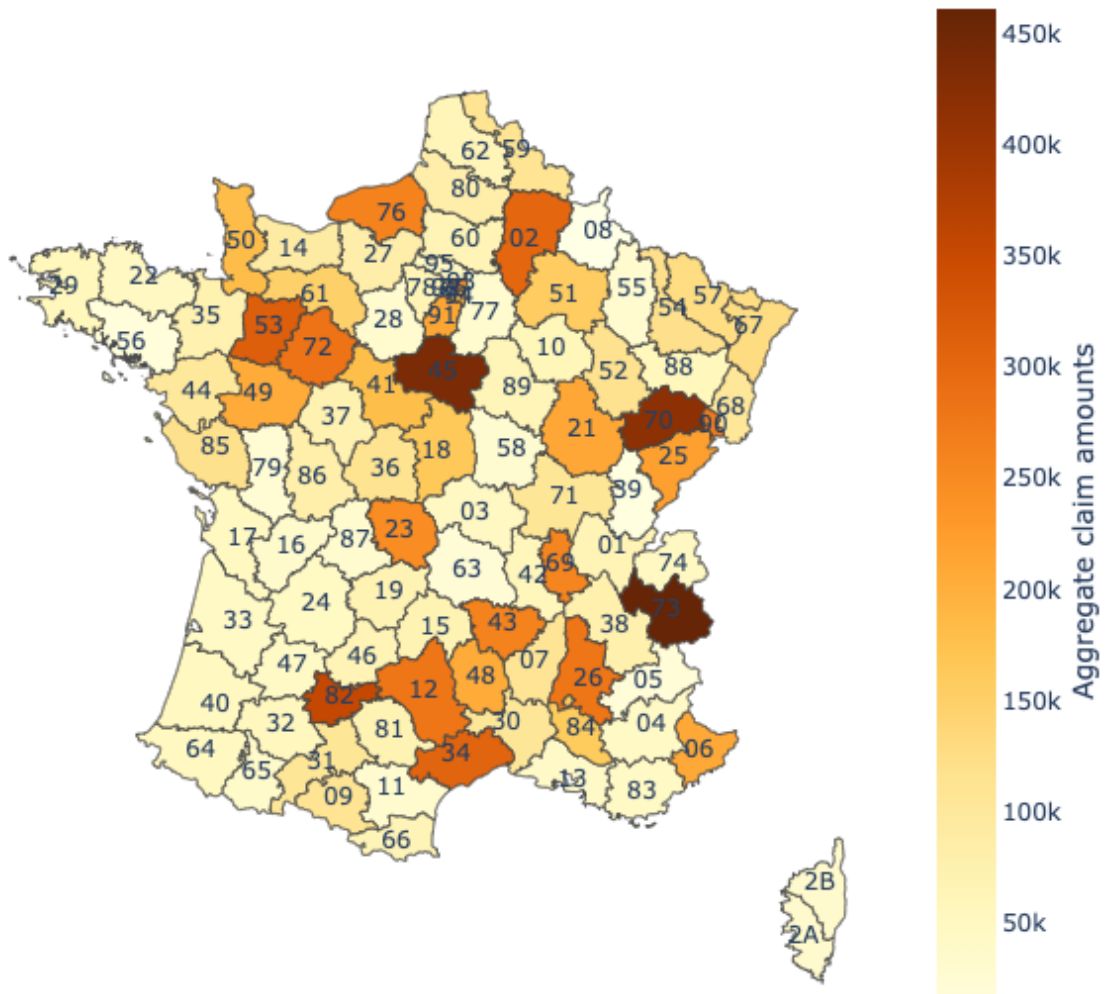}
\caption{\scriptsize The total amount of claims per departments}
\label{fig:DepAmount}
\end{figure}

\begin{figure}
\centering
\includegraphics[scale = 0.45]{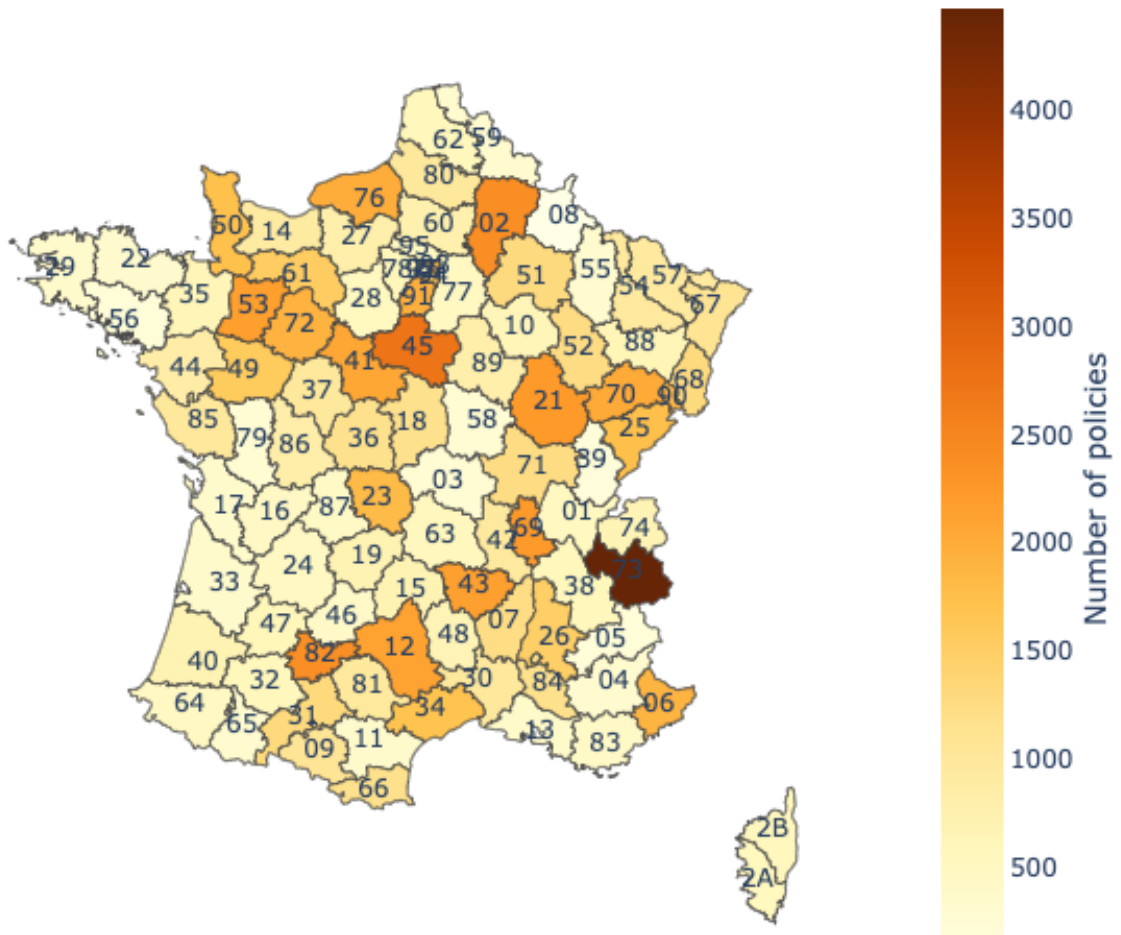}
\caption{\scriptsize The total number of policies per departments}
\label{fig:DepPolicy}
\end{figure}

\begin{figure}
\centering
\includegraphics[scale = 0.45]{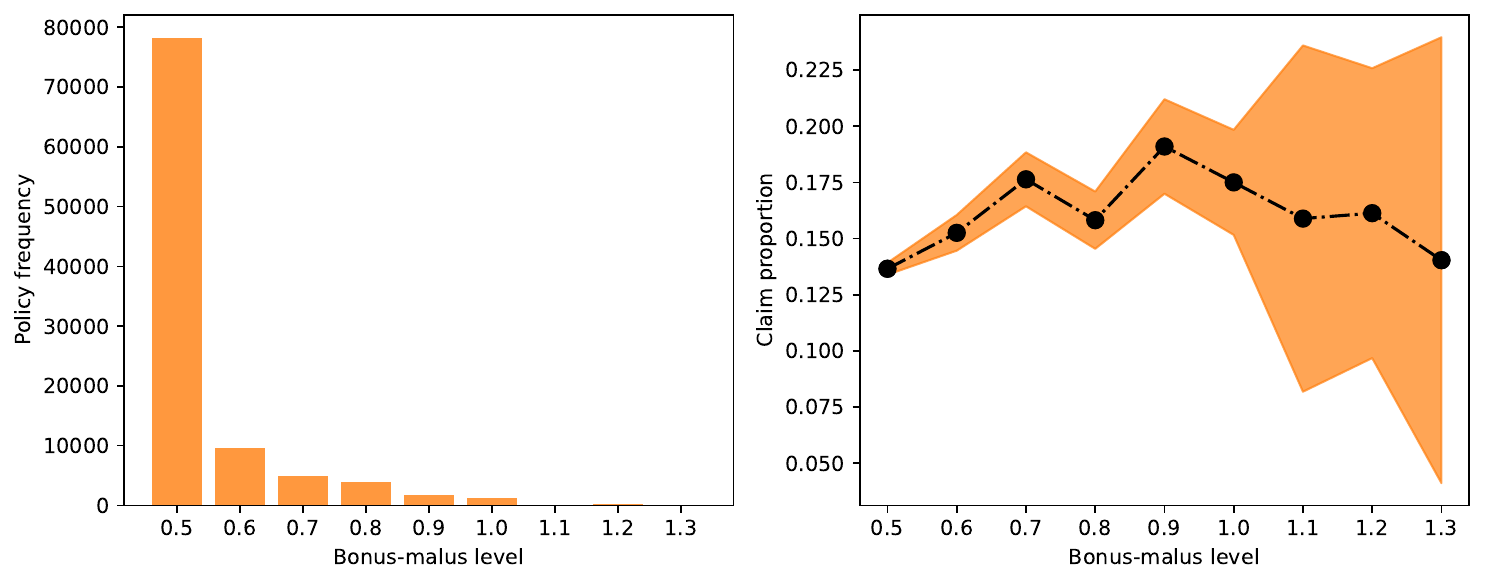}
\caption{\scriptsize The policy and claims frequency per bonus-malus levels}
\label{fig:Bonus}
\end{figure}

\begin{figure}
\centering
\includegraphics[scale = 0.45]{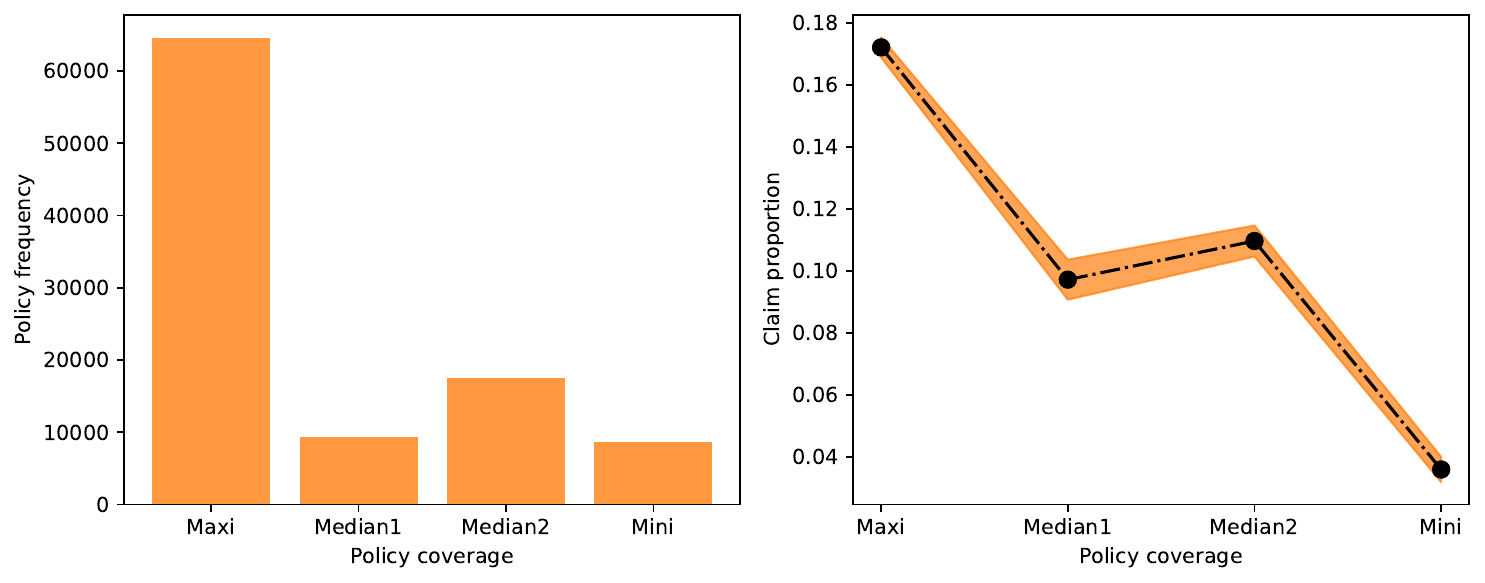}
\caption{\scriptsize The policy and claims frequency per type of coverage}
\label{fig:Coverage}
\end{figure}

\begin{figure}
\centering
\includegraphics[scale = 0.45]{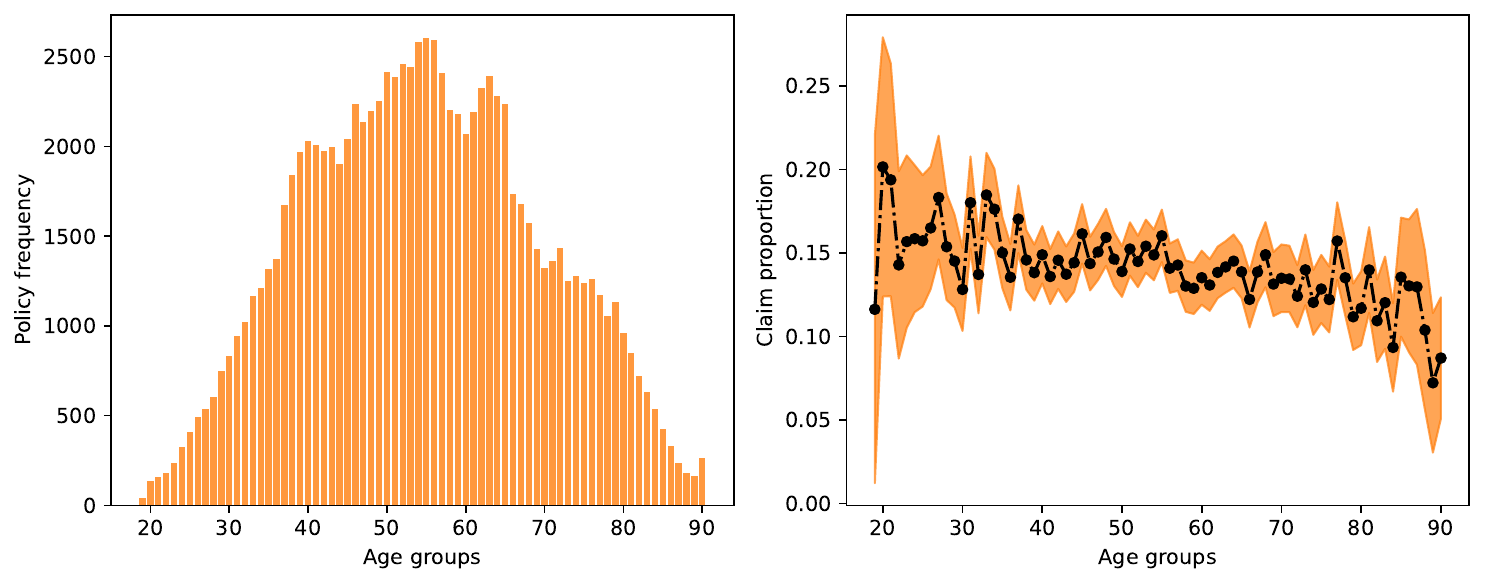}
\caption{\scriptsize The policy and claims frequency per driver age groups}
\label{fig:DrvAges}
\end{figure}

\begin{figure}
\centering
\includegraphics[scale = 0.45]{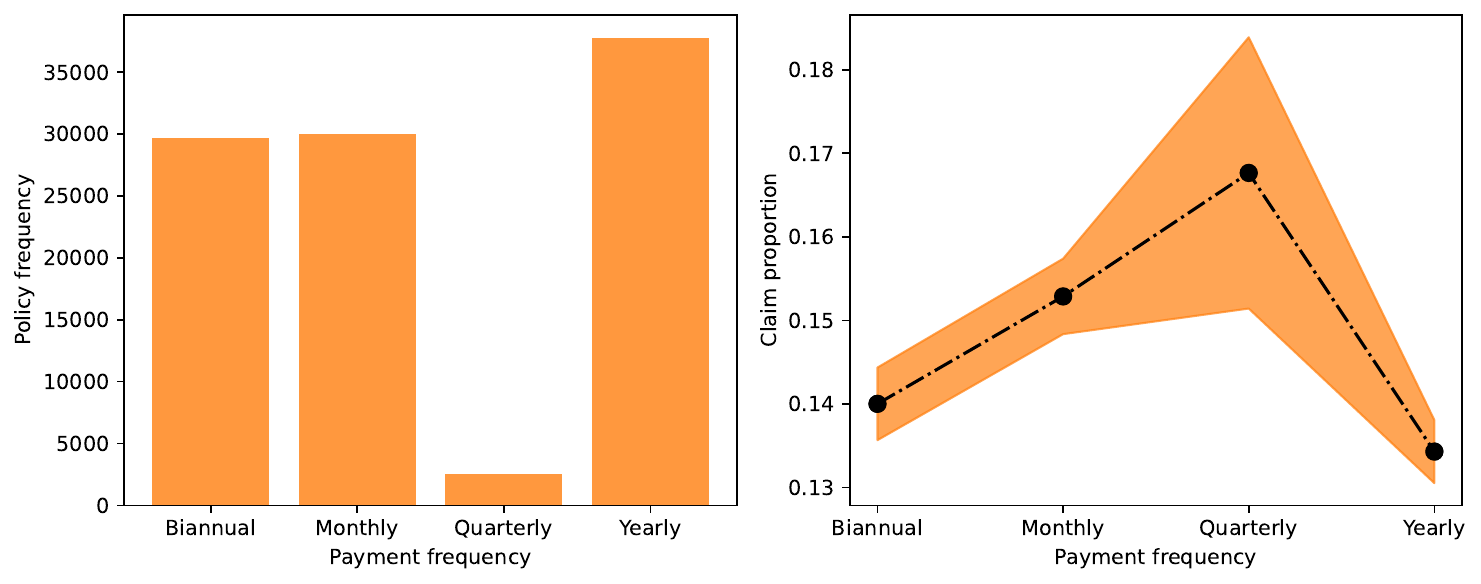}
\caption{\scriptsize The policy and claims frequency per payment frequency}
\label{fig:Frequency}
\end{figure}

\begin{figure}
\centering
\includegraphics[scale = 0.45]{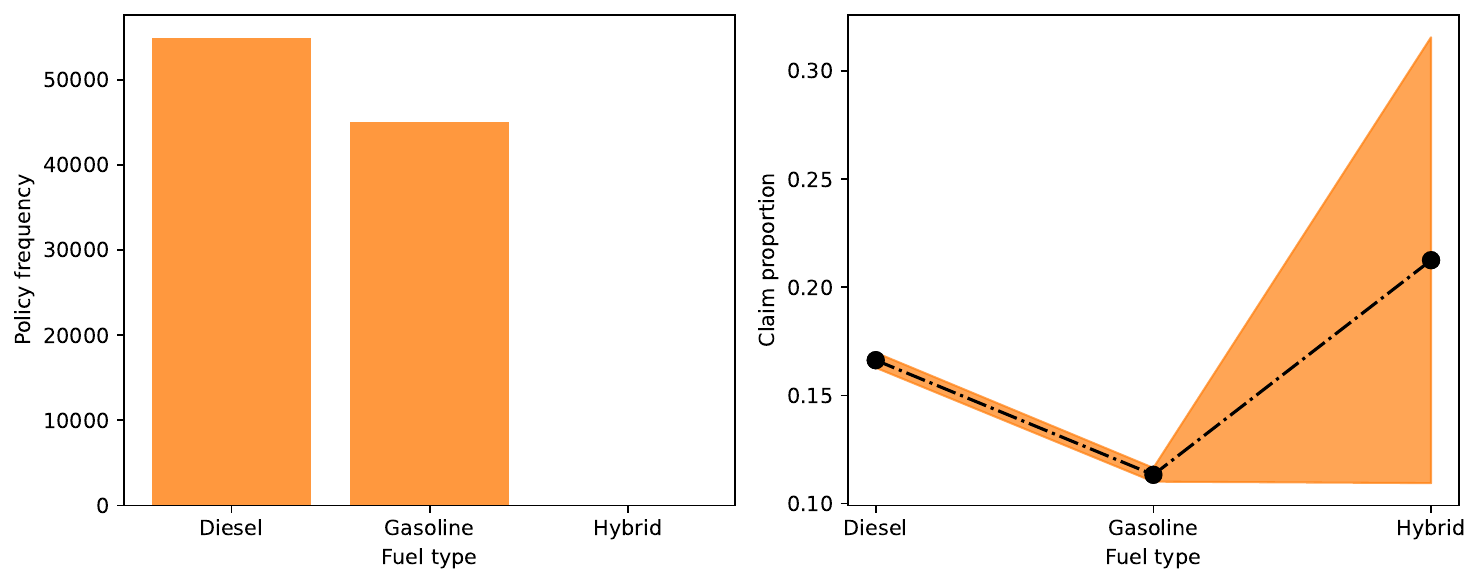}
\caption{\scriptsize The policy and claims frequency per fuel types}
\label{fig:Fuel}
\end{figure}

\begin{figure}
\centering
\includegraphics[scale = 0.45]{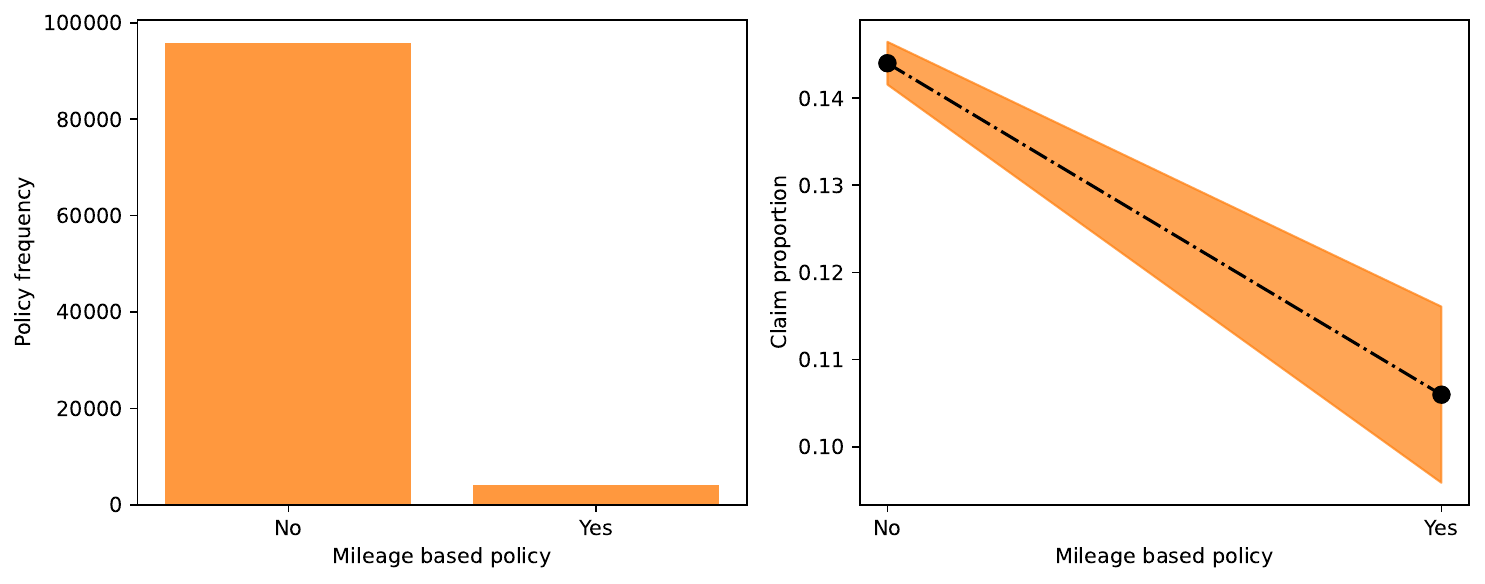}
\caption{\scriptsize The policy and claims frequency per participants in mileage-based policy}
\label{fig:Mileage}
\end{figure}

\begin{figure}
\centering
\includegraphics[scale = 0.45]{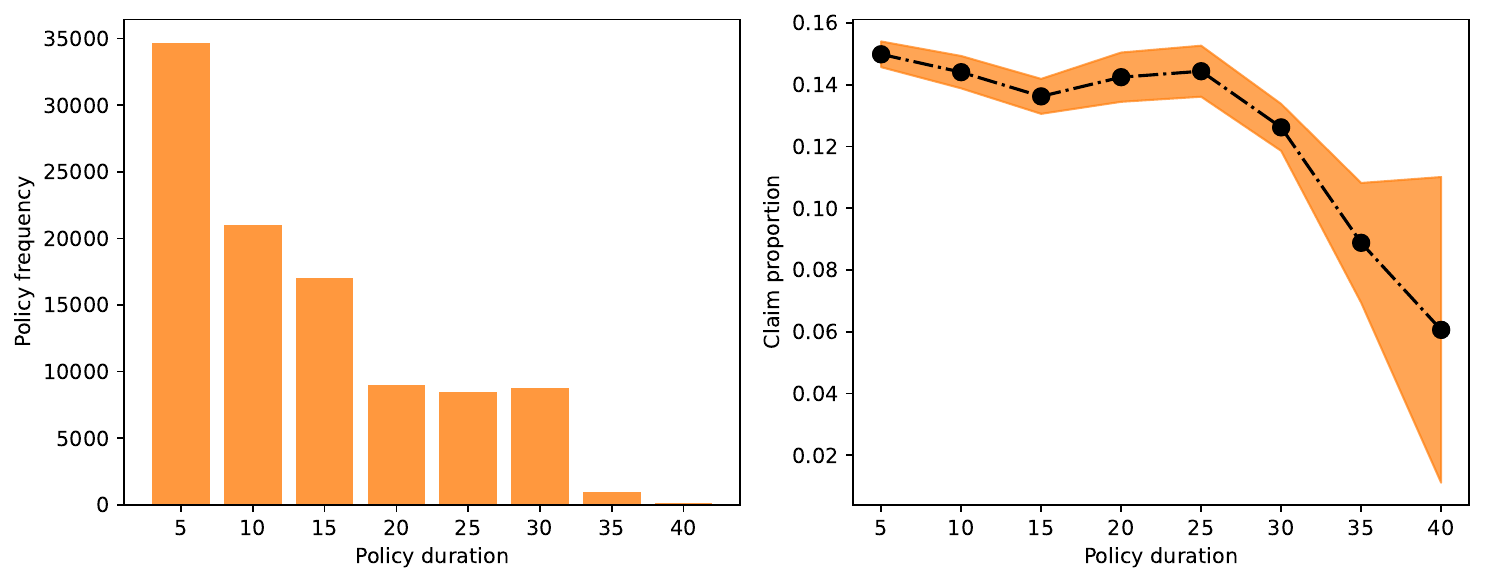}
\caption{\scriptsize The policy and claims frequency per duration}
\label{fig:Duration}
\end{figure}

\begin{figure}
\centering
\includegraphics[scale = 0.45]{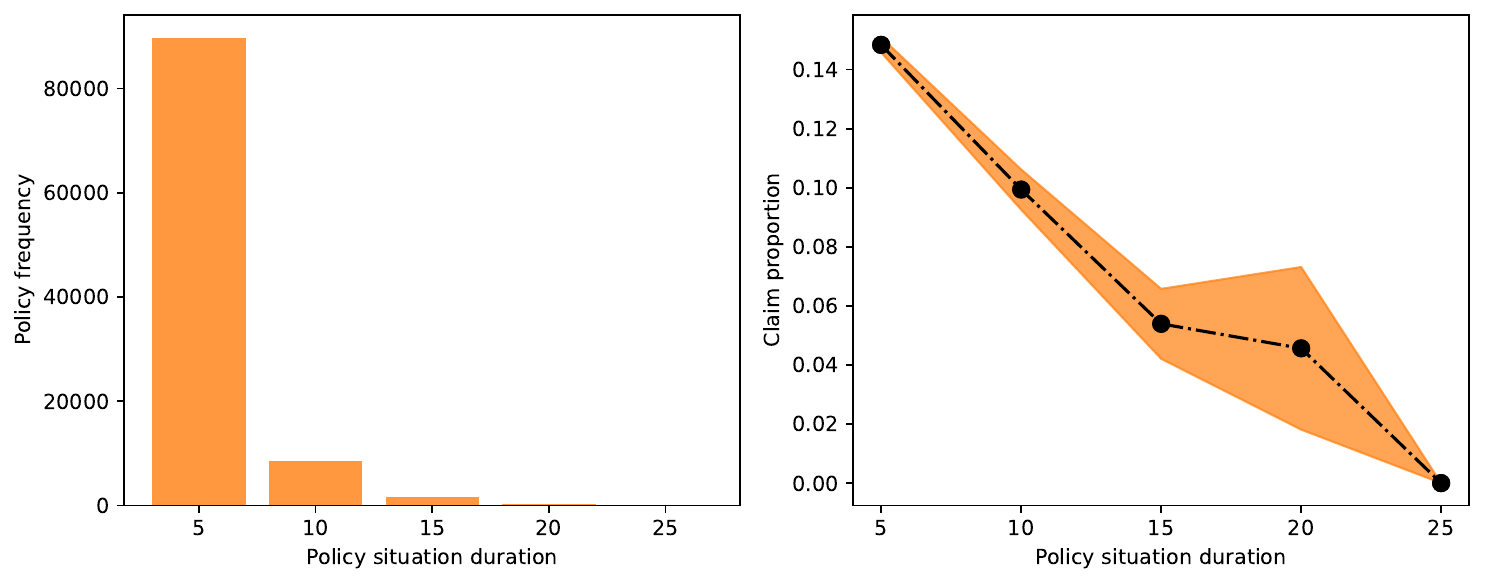}
\caption{\scriptsize The policy and claims frequency per situation duration}
\label{fig:Situation}
\end{figure}

\begin{figure}
\centering
\includegraphics[scale = 0.45]{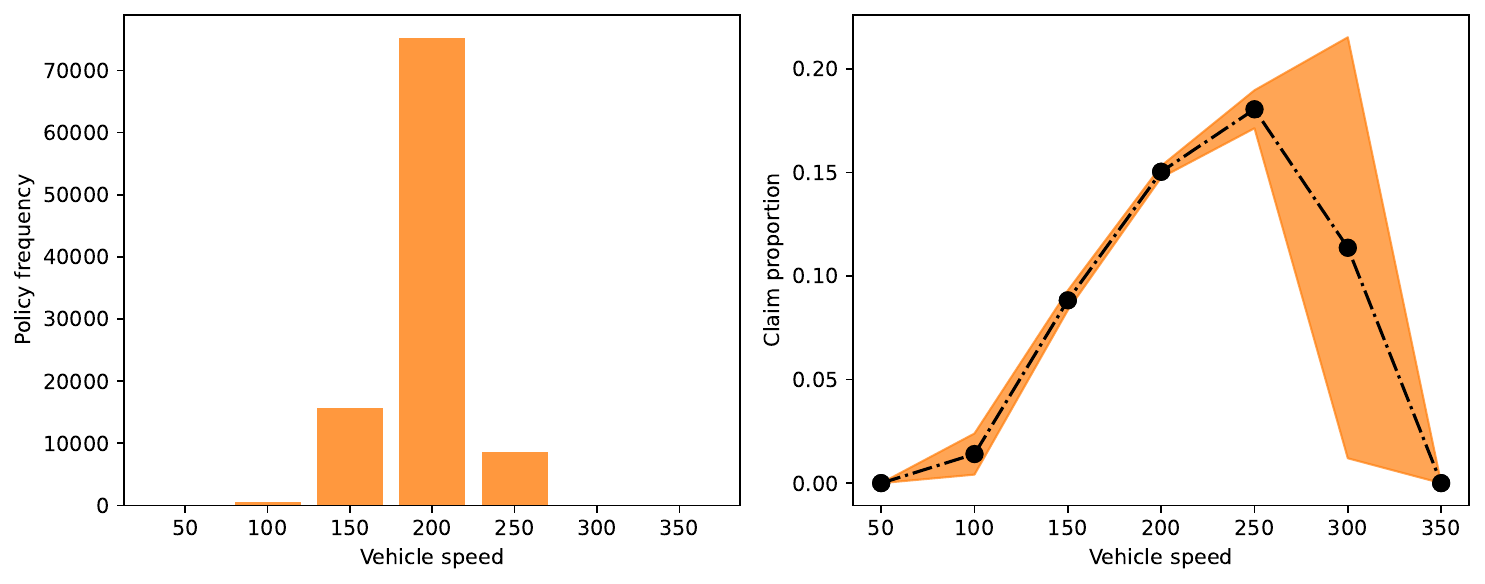}
\caption{\scriptsize The policy and claims frequency per vehicle speed}
\label{fig:Speed}
\end{figure}

\begin{figure}
\centering
\includegraphics[scale = 0.45]{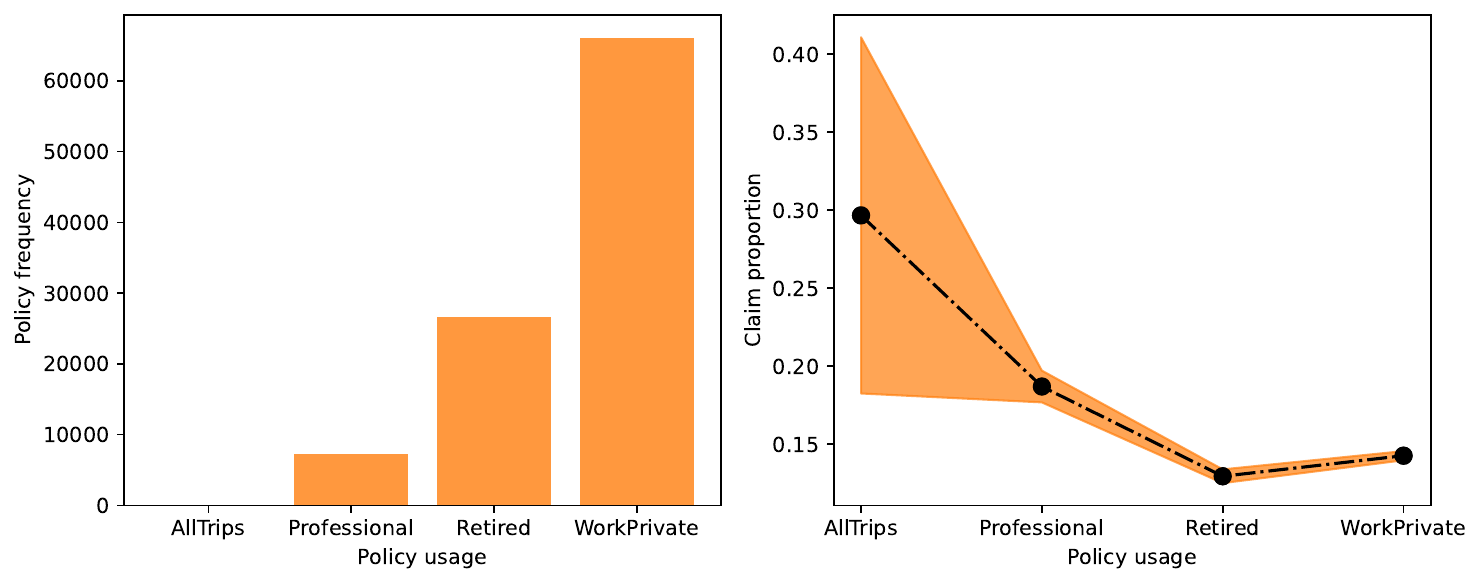}
\caption{\scriptsize The policy and claims frequency per policy usage}
\label{fig:Usage}
\end{figure}

\begin{figure}
\centering
\includegraphics[scale = 0.45]{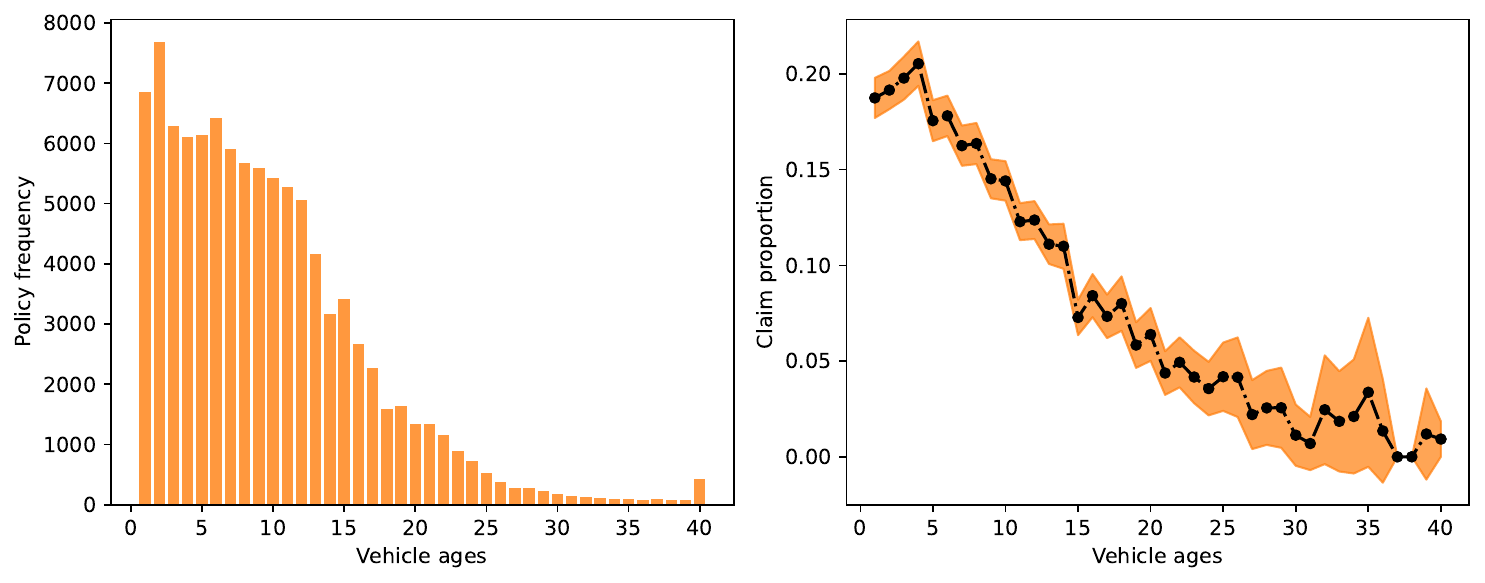}
\caption{\scriptsize The policy and claims frequency per vehicle ages groups}
\label{fig:VhAges}
\end{figure}

\begin{figure}
\centering
\includegraphics[scale = 0.45]{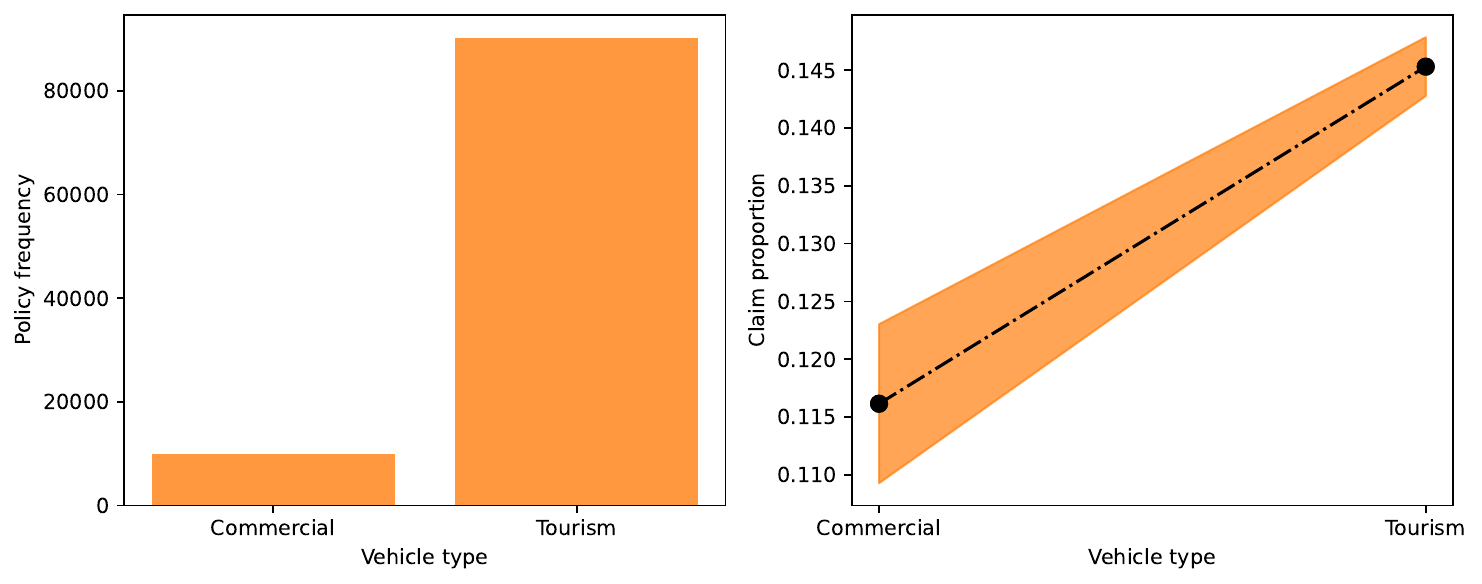}
\caption{\scriptsize The policy and claims frequency per vehicle type}
\label{fig:VhType}
\end{figure}

\begin{figure}
\centering
\includegraphics[scale = 0.45]{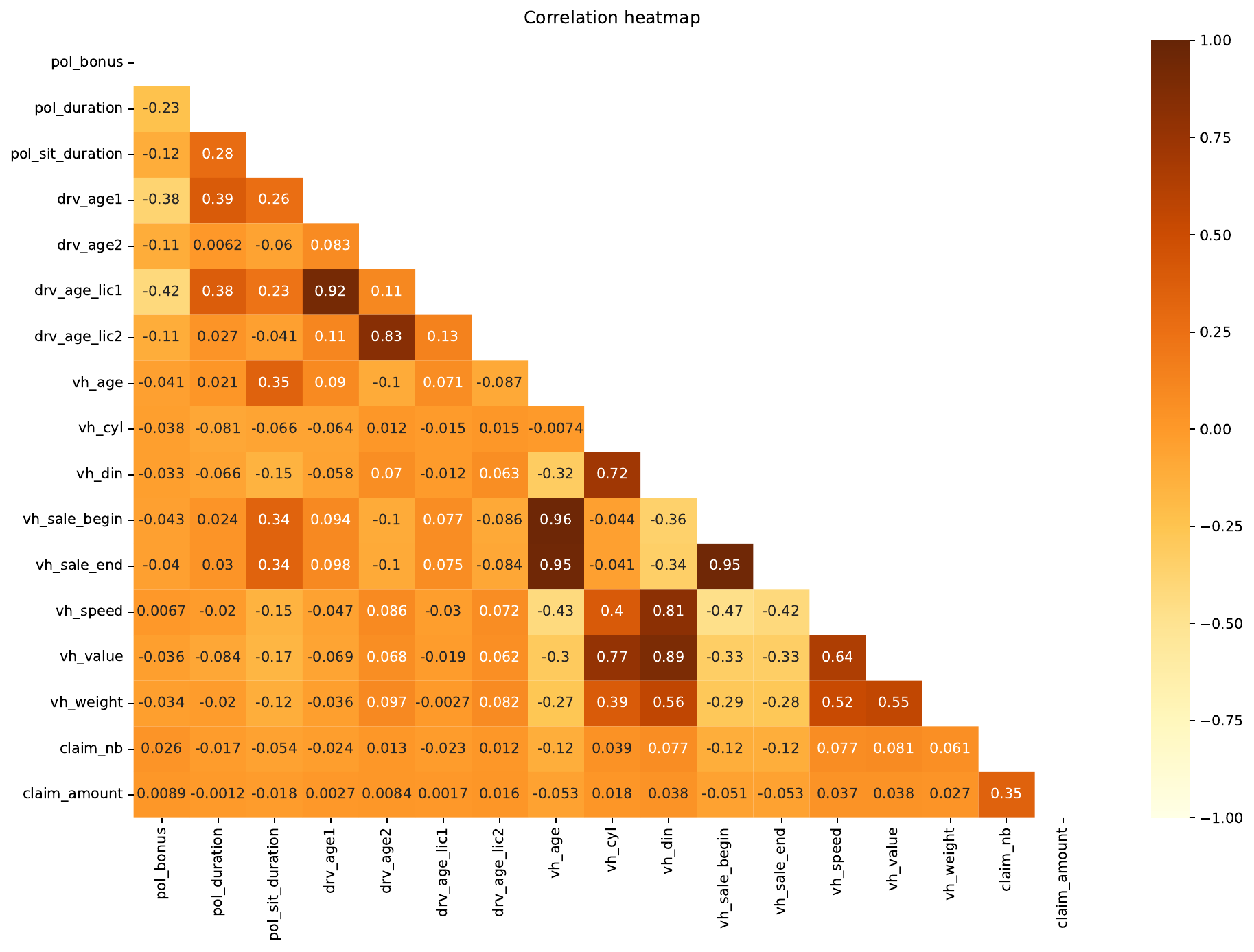}
\caption{\scriptsize The correlation heatmap}
\label{fig:Heatmap}
\end{figure}

In this study, we use the datasets used for the 2017 pricing game of the French Institute of Actuaries. Here we use datasets: $pg17trainpol$ which contains $100,000$ policies for motor third-party liability insurance (TPL) and $pg17trainclaim$ which contains $14,243$ claims of those TPL  policies. Table \ref{tab:Var} provides an overview of the features (statistical variables) and their descriptions.
Some policies cover more than $1$ car which is identified by the same $id\_client$ and different $id\_vehicle$. In $pg17trainclaim$ we combine $id\_client$ and $id\_vehicle$ to form the $id\_policy$ as in $pg17trainpol$ and then aggregate claim numbers and claim amounts for each policy. We then merge the updated claim dataset with $pg17trainpol$. After merging, the $claim\_nb$ and $claim\_amount$ for the policies without claims are shown by $NaN$. We then replace $NaN$ with $0$. In total, we have $14,243$ claims that amount to \euro{$11,724,608.37$}. There is only one missing value in $vh\_age$ which belongs to policy $A00000765$ and can be filled in by referring to $pg17testyear1$ file. Table \ref{tab:ClaimPerPolicy} shows the number of policies per claim frequency. According to this table, we have an unbalanced dataset with $87.3\%$ of policies without claims and $12.7\%$ with claims. Next, we look at the statistical properties of the number of policies, the number of claims and the amount of claims in France, which is shown by Table \ref{tab:ClaimPerDep}. To visualise these quantities, we plot the choropleth map of 96 departments in France. To do this, we use the first $2$ digits of $pol\_insee\_code$ \footnote{\url{http://cas.uqam.ca/pub/web/CASdatasets-manual.pdf}} and $geojason$ file of 96 departments of France that can be obtained from GitHub repository \footnote{\url{https://github.com/gregoiredavid/france-geojson}}. Figure \ref{fig:Dep} illustrates the choropleth map of 96 departments in France and the numbers represent department codes. Figure \ref{fig:DepClaim} shows the aggregate number of claims for each department. As we can see the maximum number of claims belongs to Var with department code 73 and the next highest claim numbers belong to Loiret with code 45. In Figure \ref{fig:DepAmount} the same departments represent the highest amount of claims and in Figure \ref{fig:DepPolicy} Var has the maximum number of policies. After looking at the policy and the claim frequency per department, we analyse these quantities per different features. Figure \ref{fig:Bonus} on the left-hand side illustrates the number of policies for each bonus-malus levels and on the right-hand side illustrates the claim proportion, i.e. the ratio of the number of claims to the number of policies for different bonus-malus levels and $95\%$ confidence level. We can observe that most policies are at the bonus level of $0.5$ which is the lowest coefficient. Also, the highest proportion of claims per number of policies belongs to bonus level of 0.9. The confidence interval is very wide after the bonus level of 0.9 where we do not have sufficient number of policies after that level. In Figure \ref{fig:Coverage} we can see that a large number of policies are assigned to $Maxi$ which provides full coverage. This type of policy also produces a large number of claims compared to the number of policies. As we have sufficient number of policies in each category, the confidence interval is narrow. Figure \ref{fig:DrvAges} shows a large number of policies belonging to ages between 40 and 60 and as we can see the younger the policyholders, the larger is the proportion of claims which satisfies our expectation as young drivers have little driving experience. From Figure \ref{fig:Frequency} we can see that very few policies pay premiums quarterly and the number of claims compared with the volume of policies is high for this category. Figure \ref{fig:Fuel} shows that most insured cars use $diesel$ and $gasoline$. There are very few cars with hybrid fuels which are not observable here, but from the right figure, we can see that most of those cars produce claims. Figure \ref{fig:Mileage} shows that most policyholders do not agree to sign up for the mileage-based policy. According to Figure \ref{fig:Duration} and Figure \ref{fig:Situation} at least 5 years have passed since the issue of most policies and at most 5 years have passed since the changes in most policies, respectively. Figure \ref{fig:Speed} shows that the maximum speed for most insured cars is 200 km/h and the proportion of claims for cars with the maximum speed of 250 km/h is higher than other cars. Figure \ref{fig:Usage} shows that most cars are being used for $private$ purposes, but the proportion of claims for these cars is less than $professional$ usage. Figure \ref{fig:VhAges} shows that most cars are younger than 10 years and the proportion of claims for younger cars is more than older cars. Figure \ref{fig:VhType} shows that most cars are used as $tourism$ and the proportion of claims for this category is higher than $commercial$ category. Figure \ref{fig:Heatmap} displays the correlation heatmap between continuous variables in Table \ref{tab:Var}. We can observe a strong correlation between driver ages and the age of obtaining a driving license, the age of the vehicle and the sale price of the car from the beginning to the end of marketing years, the properties of the car, such as speed, cylinder, motor power, weight and its value.

\begin{table}
\scriptsize
\caption{An overview of the features used in our study}
\label{tab:Var}
\begin{tabular}{lll}\toprule
id\_policy	&  	& unique identification number.\\
pol\_bonus &  French no-claim discount system & It starts at $1$. Every year without claim, the bonus coefficient decreases\\
 && by $5$ percent until it reaches $0.5$. Every time the driver causes a claim, \\
&& the coefficient increases by $0.25$ with a maximum of $3.5$.\\\\
pol\_coverage & Coverage category & $4$ types: Mini (it only covers Third Party Liability claims), \\
&& Median1, Median2, Maxi (it covers all claims).\\\\
pol\_duration & Policy duration & It represents how old the policy is. \\
pol\_sit\_duration & Situation duration & How old the current policy characteristics are. Policies can be evolved, \\
&&e.g. by changing coverage, vehicle, drivers, etc.\\\\
pol\_pay\_freq & The payment frequency & Annually, bi-annually, quarterly, monthly.\\
pol\_payd & Mileage-based policy & Yes indicates that the driver has subscribed to a mileage-based policy.\\
pol\_usage & Policy usage & 4 categories: ``WorkPrivate", ``Retired", ``Professional", ``AllTrips".\\
pol\_insee\_code & INSEE code & code of the French city/municipality where the policyholder lives. \\
&&It identifies ``communes" and ``departments" in France.\\\\
drv\_drv2 &  & Secondary driver.\\
drv\_age1,\_age2 &  & The age of the ith driver.\\
drv\_sex1,\_sex2 & & The gender of the ith driver.\\
drv\_age\_lic1,\_lic2 & & The age of the driving license of the ith driver.\\\\
vh\_age & The vehicle age & The vehicle's age is the difference between the year of release\\
&& and the current year.\\
vh\_cyl & & The engine cylinder in ml.\\
vh\_din & & It is a representation of the motor power.\\
vh\_fuel & Fuel type & 3 types: ``Diesel", ``Gasoline", ``Hybrid".\\
vh\_make & The vehicle carmaker & such as Renault, Peugeot, Citroen, etc.\\
vh\_model & The vehicle model& A subdivision of the carmake.\\
vh\_sale\_begin, \_end & & The beginning and the end of marketing years of the vehicle. \\
vh\_speed & & The vehicle maximum speed (km/h) as stated by the manufacturer.\\
vh\_type & & 2 types: ``Tourism" and ``Commercial".\\
vh\_value & & The vehicle's value (replacement value) in euros.\\
vh\_weight & & The vehicle weight (kg).\\\\
claim\_nb & Aggregate claim number & Total claim number for one driver.\\
claim\_amount & Aggregate claim amount & Total claim amount for one driver ranging from $-2,000$ to $+300,000$.\\
&& Negative claims happen due to subrogation rights of the insurer.\\\bottomrule
\end{tabular}
\end{table}

\begin{table}
\scriptsize
\caption{The number of policies for different claim frequency}
\centering
\label{tab:ClaimPerPolicy}
\begin{tabular}{ccccccccc}\toprule
Claim frequency & 0 		     & 1 		& 2 		& 3 		& 4 	     & 5 & 6\\\midrule
Policy frequency & $87,346$ & $11,238$	& $1,264$ & $134$	& $16$  & 1 & 1\\\bottomrule
\end{tabular}
\end{table}

\begin{table}
\scriptsize
\caption{Statistical properties of policies in France}
\centering
\label{tab:ClaimPerDep}
\begin{tabular}{ccccccccc}\toprule
			 & Mean	      & Std Dev.   & Min 	     & 1st Quantile  & Median  & 3rd Quantile & Max\\\midrule
Claim number & $148.36$      & $115.10$	         & $3.00$       & $62.75$                       & $111.00$                      &  $216.00$                    & $611.00$\\
Claim amount & $122,131.34$& $103,003.19$  & $113.56$   & $49,427.15$		   & $76,696.35$		&   $180,539.11$  	      & $461,168.09$\\
Policy 	       & $1,041.67$	&$737.47$	& $21.00$	     & $512.50$			   & $818.00$			&  $1,511.75$			& $4,473.00$\\\bottomrule
\end{tabular}
\end{table}

\section{Classification algorithms}
Suppose we have an input vector $\mathbf{x}_i = (x_{i1}, \dots, x_{ip})^{T}$ of dimension $p$, for $i=1,\dots,n$, where the vector $\mathbf{x}_i$ contains the $p$ variables representative of the $i$th policyholder and the corresponding output are denoted by $\mathbf{y}_i$. In a classification problem $\mathbf{y}_i$ represent different classes. In this paper, we consider a binary classification problem with two classes, $y_{i1} = 1$, i.e. a policy with claims and $y_{i2} = 0$, i.e. a policy without claims. In this section, we investigate different classification algorithms and models. For more details see, for example, Lantz (2015), Müller and Guido (2016), Hastie et al. (2017) and Denuit et al. (2019a, 2019b, 2020).

\subsection{Nearest neighbour classifier}
K-nearest neighbour (KNN) algorithm is introduced by Fix and Hodges in 1951 and published in Fix and Hodges (1989). It can be used both in supervised and unsupervised machine learning problems. It classifies the target variables by looking at the relationship between features. The KNN algorithm is used in computer vision applications such as character and facial recognition, prediction of individuals' preferences and pattern recognition in genetic data such as detecting proteins or diseases. The KNN uses information about k-nearest neighbours of an instance to classify that instance. In our case, it uses information such as type of coverage, payment frequency, driver's age, etc about k-nearest policies to predict whether a claim occurs under a particular policy or not by looking at the k policies which are the ``nearest" in similarity. We can think of a policy as a point in a multidimensional space where each dimension represents one of the features. We can then classify these points by putting neighbours into one class. In order to classify one point as $1$ or $0$, we need to measure its distance from $k$ neighbours. The smaller $k$, the more complex and specific the model is. Large $k$ provides a smooth and more general algorithm. The distance between instances is defined by    
\begin{eqnarray}
\label{eq:dis}
D(\mathbf{x},\mathbf{y}) = \left(\sum_{i=1}^n |x_i-y_i|^p)\right)^{1/p}
\end{eqnarray}
where $x_i$ denotes the $i$th feature of policy $x$ and $y_i$ denotes the $i$th feature of policy $y$. Equation \ref{eq:dis} is known as Minkowski distance and is reduced to ``euclidean" distance for $p=2$ and ``manhattan" distance for $p=1$. Other distance functions include ``chebyshev" and ``mahalanobis" distances. One of the advantages of this algorithm is that it is simple, fast and does not make any assumptions about the underlying data distribution. To implement this algorithm first we need to process our data. The KNN algorithm cannot be used with non-numeric data and we need to convert the nominal and ordinal features into numeric values.

\subsection{Logistic regression model}
The logistic regression model is classified under linear models. A simple linear model is a model that describes a linear relationship between different features and is given by
\begin{eqnarray}
\label{eq:linmod}
y_i(\mathbf{x}) = w_0 + \sum_{j=1}^p w_j x_{ij} =  \mathbf{x}_i^T \mathbf{w}, \qquad i = 1, 2, \dots, n,
\end{eqnarray}
where $\mathbf{w} = (w_0, w_1, \dots, w_p)^T$ is a vector of dimension $p+1$ containing the unknown regression coefficients. 
In a classification problem, we need to constrain the value of $y$ in the range $(0,1)$ to treat $y$ as the probability that an object falls into a particular class. For this we use Generalised Linear Models (GLMs). Let $f(.)$ be a nonlinear function which is known as an ``activation function" in ML and its inverse $g^{-1}(.)$ is a ``link function" in statistics. We then rewrite \ref{eq:linmod} as
\begin{eqnarray}
f\left( \mathbf{x}_i\right) = g^{-1}\left( w_0 + \sum_{j=1}^p w_j x_{ij} \right). \nonumber
\end{eqnarray}
In a logistic model, we define the activation function by a ``logistic" or ``sigmoid" function, given by
\begin{eqnarray}
f(x) = \frac{1}{1+e^{-x}}\nonumber.
\end{eqnarray} 
To find the coefficients $\mathbf{w}$ we need to solve an optimisation problem. We can do this by minimising the objective function or maximising the likelihood function. To avoid overfitting, we can also add a penalty term and control the regularisation parameter. 
\begin{eqnarray}
\label{eq:penalty}
\min_{\mathbf{w}} \frac{1}{2} \sum_{i=1}^n (\hat{y}_i - y_i)^2 + \frac{\lambda}{2} \sum_{j=1}^p |w_j|^k
\end{eqnarray}
where $\lambda$ is a tuning parameter. As $\lambda$ increases, more coefficient estimates are set to 0. In equation \ref{eq:penalty} $k=1$ corresponds to the Least Absolute Selection and Shrinkage Operator (LASSO) and $k=2$ corresponds to ridge penalties. Further, combining LASSO and ridge penalties yields an elastic net regularisation.

\section{Results}
In this section, we compare KNN and logistic regression with different regularisation parameters.

\begin{verbatim}
X_d = pd.get_dummies(data = X, columns=['pol_coverage','pol_pay_freq',
'pol_payd','pol_usage','drv_drv2','drv_sex1','vh_fuel','vh_type'])
\end{verbatim}
 This produces one column for each category in a feature. For example, an $n$-category feature produces $n-1$ new features in which a value of $1$ indicates that category and $0$ otherwise. Then we need to rescale the features as their range can be quite different. We can use min-max normalisation or z-score standardisation, which can be achieved by
\begin{verbatim}
X_s = StandardScaler()
X_d = pd.DataFrame(X_s.fit_transform(X_d),columns = X_d.columns)
\end{verbatim}
When we call the function for training k-NN in ``sklearn", we need to determine the metric and the weights. Here, we choose the inverse of the distance between the points as the weights in order to put more emphasis on the closer points than those which are further away. We also need to determine an appropriate value of k. From Figure \ref{fig:Accuracy} we can see that the accuracy level for both training and test sets is about $0.87$ when k$=20$. To evaluate our model, we split our data into training and test sets. The codes are as follows
\begin{verbatim}
X_d_train, X_d_test, y_train, y_test = train_test_split(X_d, y, random_state=0)
model = KNeighborsClassifier(algorithm='auto', leaf_size=30, metric='minkowski', 
metric_params=None, n_jobs=1,n_neighbors=20, p=2, weights='distance')
model.fit(X_d_train, y_train)
y_predict = model.predict(X_d_test)
model.score(X_d_test,y_test)
\end{verbatim}

\begin{figure}
\centering
\includegraphics[scale = 0.7]{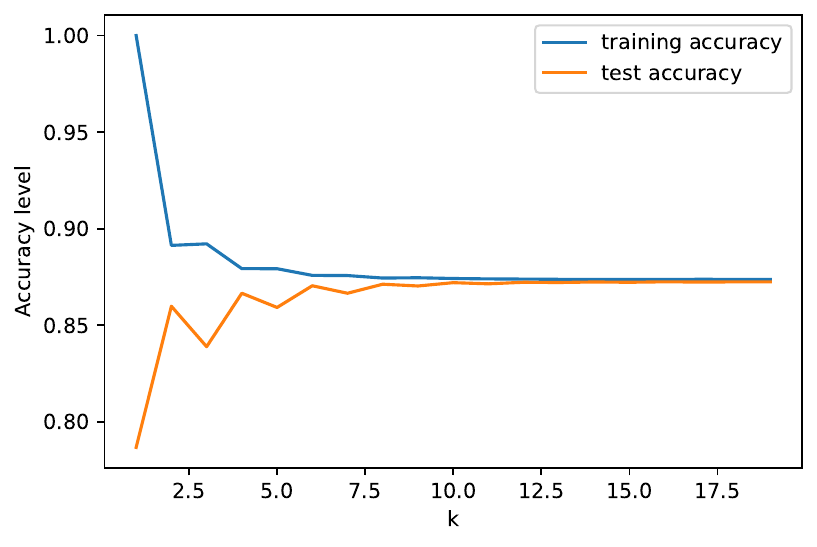}
\caption{ Training and test accuracy for different values of k}
\label{fig:Accuracy}
\end{figure}

\begin{figure}
\centering
\includegraphics[scale = 0.7]{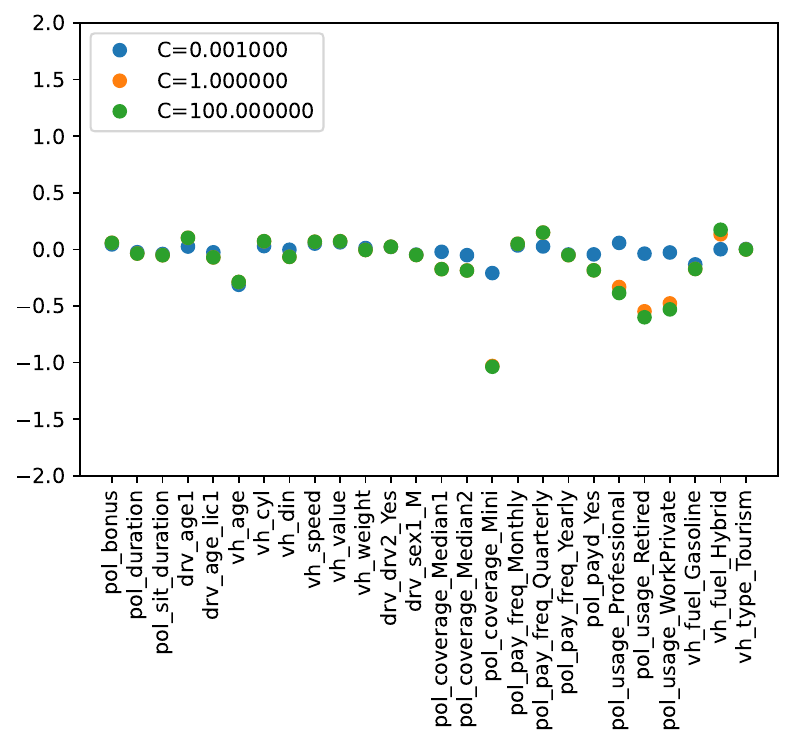}
\caption{ Parameters for different values of C (regulalisation parameter)}
\label{fig:Accuracy}
\end{figure}

\subsection{Confusion matrix}
To evaluate a classification model we can use $N \times N$ classification matrix, where $N$ is the number of classes. In our case, we have $2 \times 2$ confusion matrix which compares the predicted values with actual ones and is given by Table \ref{tab:Conf}. We can see that this algorithm predicts $21,804$ cases ``without claims" and $0$ cases ``with claims" accurately. We can change True Positive (TP) and True Negative (TN) by changing k and the other parameters such as ``weights". For example, with k$=3$ and uniform weights, the algorithm can predict $208$ cases ``with claims" and $20,764$ cases ``without claims" accurately. We can calculate ``precision" and ``recall" level from confusion matrix:
\begin{eqnarray}
\textnormal{Precision}: \frac{\textnormal{TP}}{\textnormal{TP}+\textnormal{FP}} = 0.87\nonumber\\
\textnormal{Recall}: \frac{\textnormal{TP}}{\textnormal{TP}+\textnormal{FN}} = 0.99\nonumber
\end{eqnarray}

\begin{table}
%\scriptsize
\caption{Confusion matrix for k-nearest neighbour algorithm}
\centering
\label{tab:Conf}
\begin{tabular}{cc|cc}\toprule
	&	& \multicolumn{2}{c}{Predicted values}\\\midrule
\multicolumn{1}{c|}{Actual}	&	      		       &     Without claims 	    & With claims\\
\multicolumn{1}{c|}{values}	&      Without claims & TP:$21,804$		    &	FN:$8$	\\	
\multicolumn{1}{c|}{}			&      With Claims     & FP:$3,188$		    &  TN: $0$\\\bottomrule
\end{tabular}
\end{table}

\begin{table}
\caption{Confusion matrix for logistic regression model}
\centering
\label{tab:Conf}
\begin{tabular}{cc|cc}\toprule
	&	& \multicolumn{2}{c}{Predicted values}\\\midrule
\multicolumn{1}{c|}{Actual}	&	      		       &     Without claims 	    & With claims\\
\multicolumn{1}{c|}{values}	&      Without claims & TP:$21,837$		    &	FN:$0$	\\	
\multicolumn{1}{c|}{}			&      With Claims     & FP:$3,163$		    &  TN: $0$\\\bottomrule
\end{tabular}
\end{table}

\break


\begin{thebibliography}{999}
\bibitem{} Breiman, L. Friedman, J.H., Olshen, R.A. and Stone, C.J. (1984). Classification and regression trees. Chapman and Hall/CRC.
\bibitem{} Campbell, M. (1986). An integrated system for estimating the risk premium of individual car models in motor insurance. \textit{ASTIN Bulletin}, 16(2), 165--183.
\bibitem{} Carfora, M.F., Martinelli, F., Mercaldo, F., Nardone, V., Orlando, A., Santone, A. and Vaglini, G. (2019). A ``pay-how-you-drive" car insurance approach through cluster analysis. \textit{Soft Computing}, 23(9), 2863--2875.
\bibitem{} Denuit, M., Hainaut, D. and Trufin, J. (2019a). Effective statistical learning methods for actuaries I: GLMs and extensions. Springer.
\bibitem{} Denuit, M., Hainaut, D. and Trufin, J. (2019b). Effective statistical learning methods for actuaries III: neural networks and extensions. Springer.
\bibitem{} Denuit, M., Hainaut, D. and Trufin, J. (2020). Effective statistical learning methods for actuaries II: tree-based methods and extensions. Springer.
\bibitem{} Denuit, M., Mesfioui, M. and Trufin, J. (2019a). Bounds on concordance-based validation statistics in regression models for binary responses. \textit{Methodology and Computing in Applied Probability}, 21, 491--509. 
\bibitem{} Denuit, M., Mesfioui, M. and Trufin, J. (2019b). Concordance-based predictive measures in regression models for discrete responses. \textit{Scandinavian Actuarial Journal}, 2019(10), 824--836.
\bibitem{} Devriendt, S., Antonio, K., Reynkens, T. and Verbelen, R. (2021). Sparse regression with multi-type regularised feature modeling. \textit{Insurance: Mathematics and Economics}, 96, 248--261.
\bibitem{} Dutang, C., Charpentier, A. and Dutang, M.C. (2020). Package 'CASdatasets'.
\bibitem{} Fix, E., Hodges, J.L. (1989). Discriminatory analysis. Nonparametric discrimination: consistency properties. \textit{International Statistical Review}, 57(3), 238--247.
\bibitem{} Gan, G. and Valdez, E.A. (2020). Data clustering with actuarial application. \textit{North American Actuarial Journal}, 24(2), 168--186.
\bibitem{} Gao, G., Meng, S. and W\"{u}thrich, M.V. (2019). Claims frequency modeling using telematics car driving data. \textit{Scandinavian Actuarial Journal}, 2019(2), 143--162.
\bibitem{} Hastie, T., Tibshirani, R. and Friedman, J. (2017). The elements of statistical learning. Data mining, inference, and prediction. Springer.
\bibitem{} Henckaerts, R., C\^{o}t\'{e}, M.P., Antonio, K. and Verbelen, R. (2021). Boosting insights in insurance tariff plans with tree-based machine learning methods. \textit{North American Actuarial Journal}, 25(2), 255--285.
\bibitem{} Lantz, B. (2015). Machine learning with R, second edition. Packt Publishing, Birmingham, UK.
\bibitem{} Li, Y., Yan, C., Liu, W. and Li, M. (2018). A principle component analysis-based random forest with the potential nearest neighbor method for automobile insurance fraud identification. \textit{Applied Soft Computing}, 70, 1000--1009.
\bibitem{} Müller, A. and Guido, S. (2016). Introduction to machine learning with Python - a guide for data scientists. O'Reilly Media, Inc. CA.
\bibitem{} Pesantez-Navaez, J., Guillen, M. and Alca\~{n}iz, M. (2019). Predicting motor insurance claims using telematics data -- XGBoost versus logistic regression. \textit{Risks}, 7(2), 70.
\bibitem{} Qazvini, M. (2019). On the validation of claims with excess zeros in liability insurance: A comparative study. \textit{Risks}, 7(3), 71.
\bibitem{} W\"{u}thrich, M.V. (2018). Feature extraction from telematics car driving heatmaps. \textit{European Actuarial Journal}, 8, 383--406.




\end{thebibliography}
\end{document}